\documentclass[12pt]{iopart}
\usepackage{iopams}
\usepackage{setstack}
\usepackage{bm}  
\usepackage{graphicx}
\begin{document}

\title{Parity violation in hydrogen revisited}
\author{R. W. Dunford and R. J. Holt}
\address{Argonne National Laboratory, Argonne, IL 60439}
\ead{dunford@anl.gov}
\begin{abstract}
We reconsider
parity violation experiments 
in atomic hydrogen and deuterium
in the light of existing
tests of the Electroweak interactions, and assess whether new experiments,
using improved experimental techniques,
could make useful contributions to testing the Standard Model (SM).
We find that, 
if parity experiments in hydrogen can be done, they remain
highly desirable because 
there is negligible atomic-physics uncertainty
and low energy tests of weak
neutral current interactions
are needed to probe for new physics beyond the SM.
Of particular interest would be a measurement of the nuclear spin
independent coupling $C_{1D}$ for the deuteron at a combined error (theory +
experiment) of 0.3\%. This would provide a factor of three improvement to the
precision on $\sin^2 \theta_W$ at very low 
momentum transfer provided by heavy atom Atomic
Parity Violation (APV) experiments.  
Also, experiments in H and D could provide precise
measurements of three
other electron-nucleon,
weak-neutral-current 
coupling constants:
$C_{1p}$, $C_{2p}$,
and $C_{2D}$, which have not been accurately
determined to date. 
Analysis of a generic APV experiment in deuterium indicates 
that a 0.3\% measurement of $C_{1D}$ requires development 
of a slow (77K) metastable beam of $\approx 5\times10^{14}$D(2S)$s^{-1}$
per hyperfine component. The advent of
UV radiation from free electron laser (FEL)
technology could allow  production of such a beam.

\end{abstract}

\maketitle

\section{Introduction}
In this report, we answer two questions concerning measurement of 
atomic parity violation (APV) in hydrogen and deuterium. First, given 
the current level of tests of the Standard Model (SM), how well 
would parity violation
need to be measured in hydrogen and deuterium to make
a useful contribution to this body of data?
Second,  what experimental parameters are needed 
to achieve the
required precision? 

As to the first question, there exists a long list
of impressive experimental tests of the SM
\cite{yao06:jpg}, 
many involving large collaborations in high energy physics.
However, as discussed in section \ref{theory}, 
future progress may, in part, require additional
high-precision, low-energy measurements\cite{erl03:prd},
and in this regime, experiments in hydrogen
have the advantage that the atomic structure is known to high precision
and would not limit the theoretical interpretation. 

Another advantage of parity experiments in hydrogen is that all of the
electron-nucleon, weak-neutral-current couplings can be determined
by performing measurements at several level crossings in both
H and D. This point is discussed
in Section \ref{mixing} where we present
updated SM predictions for the parity violation mixings
at the various crossings. A summary of the results of former
PNC measurements in hydrogen is presented in
Section \ref{currentlimits}, and
in Sections \ref{sensitivity}, and \ref{systematics}
we consider our question concerning the experimental 
requirements by evaluating a generic
APV experiment in deuterium. Finally,
we summarize our conclusions
in section \ref{summary}.

\section{Theoretical Motivation\label{theory}}
Testing the SM of the electroweak interaction remains an important task of experimental physics.
This effort has reached the level of precision measurements.
For example, the weak mixing angle $\sin^2\theta _W$ has been
determined to 0.06\% at the Z boson mass. Most of the data 
confirming the SM have been obtained at high energy
near the Z resonance. 
The excellent agreement at the Z-pole was obtained by the 
weighted average of two experiments that disagree 
with one another by 3.4 $\sigma$.
Moreover, 
there have been recent 
measurements at lower energy, such as
the muon g-2 experiment
\cite{ben06:prd}, the NuTeV neutrino deep inelastic scattering experiment\cite{zel02:prl}, 
and the SLAC E-158
polarized M{\o}ller Scattering experiment \cite{ant04:prl}.
In the muon g-factor experiment there is a 3.4 standard deviation
difference between the experimental result and the theoretical
calculation\cite{mil07:rpp} suggesting new physics
beyond the Standard Model.
This situation has motivated additional
low energy measurements of the electroweak interactions, 
such as the Qweak \cite{car01:jla,grs05:epja} and DIS-Parity
\cite{rei04:aip,arr05:jla} experiments at the
Thomas Jefferson National Accelerator Facility (JLab). 

A useful way to assess the low energy tests of the electroweak interaction
is to consider the predicted running
of weak mixing angle $\sin^2\theta_W$. The SM predicts that
this parameter evolves as a function
of momentum transfer \cite{erl03:prd,fer04:epj,erl05:prd}.
This evolution can be checked
by performing precision low energy experiments 
and comparing the results with precise measurements
at high energy. In this respect, tests done near zero momentum
transfer such as APV or low energy 
electron scattering experiments are ideal.
In the $\overline{MS}$ scheme,
Erler and Ramsey-Musolf \cite{erl03:prd,erl05:prd} find for the value of
the weak mixing angle at zero momentum transfer:
\begin{equation}
\sin^2 \hat{\theta}_W (0)=0.23867 \pm 0.00016.
\label{weaklow}
\end{equation}
In the same scheme the value at the Z-pole is \cite{yao06:jpg}:
\begin{equation}
\sin^2 \hat{\theta}_W (M_Z)=0.23122 \pm 0.00015,
\label{weakhigh}
\end{equation}
so that the zero momentum transfer value is larger by about 3\%. 
Several low energy measurements test this prediction.
APV (Cs\cite{woo97:sci} 
and Tl\cite{edw95:prl,vet95:prl}) establishes the running of weak
mixing angle with a 
0.9\% measurement of $\sin^2 \hat{\theta}_W$
that is one standard deviation below
the SM prediction.
The E-158 M{\o}ller scattering experiment has an error
of 0.6\% which is one sigma above the SM.  NuTeV, a deep
inelastic neutrino scattering experiment, reports a
3$\sigma$ discrepancy with the SM value.

The Qweak experiment aims to improve the limits on the semi-leptonic
weak neutral current couplings by measuring the weak charge
of the proton Q$_W$(p). 
In lowest order the weak charge is given by\cite{bou74:pl}:
\begin{equation}
Q_W(p)=(1- 4\sin^2 \theta_W).
\label{qweak}
\end{equation}
Since $\sin^2\theta_W$ is close to 1/4, $Q_W(p)$ is  suppressed
and particularly
sensitive to higher order corrections and new physics beyond the SM.

To be more precise, we can use the SM expressions for the 
electron-hadron processes from 
Ref.~\cite{yao06:jpg}, including radiative corrections,
to derive the relation between the uncertainties in the weak charge
$Q_W(Z,N)$ and $\sin^2 \hat{\theta}_W$
for an atom with nuclear charge $Z$ and
neutron number $N$. We find:
\begin{equation}
\frac{\Delta(\sin^2 \theta_W)}{\sin^2 \theta_W}=
\left(-0.078+1.0785\frac{N}{Z}\right)\frac{\Delta Q_W}{Q_W}.
\label{qfactor}
\end{equation}
For H, D, and Cs the factors multiplying 
$\Delta Q_W/Q_W$ are -0.078, 1.00, and 1.45, respectively.
For example, a measurement of $Q_W(p)$ with a precision of 4\%
determines $\sin^2 \theta_W$ to 0.3\%. For the Boulder Cs experiment,
if the experimental error ($\sim 0.35\%$) \cite{woo97:sci} 
and atomic theory error ($\sim 0.5 \%$) are combined in 
quadrature, the total uncertainty in 
$Q_W(Cs)$ is $\sim 0.6\%$ which corresponds to
an uncertainty of $\sim 0.9\%$ in $\sin^2 \theta_W$ based
on Eq.~\ref{qfactor}, and this
is the dominant contribution to the 
current APV limit on $\sin^2 \theta_W$. For deuterium,
Eq.~\ref{qfactor} indicates that the fractional uncertainty in 
$Q_W(D)$ is equal to the fractional uncertainty in
$\sin^2 \theta_W$.

The need for precision measurements of semileptonic weak neutral current couplings at low
momentum transfer obliges a renewed consideration of measurements in hydrogen and deuterium
where precision is not limited by calculational uncertainties. In this respect, both
the heavy atom APV experiments and polarized electron scattering
experiments are susceptible to hadronic or atomic effects.
Qweak aims to measure the  parity violating
elastic electron-proton scattering asymmetry at $Q^2$ $\approx$ 0.03 GeV$^2$. This
process is proportional to the sum of $Q_W$(p) and a proton form factor $F^p$($Q^2,\theta$).
The form factor contribution must be subtracted from the scattering asymmetry
in order to extract $Q_W$(p). In principle, this extrapolation would require a hadronic
structure calculation, but an extensive series of parity violating electron-proton scattering
experiments at higher $Q^2$ will allow extrapolation to $Q^2$=0 with an uncertainty
of about 2\% in the weak charge\cite{erl03:prd}. 
The heavy atom APV experiments require sophisticated atomic theory calculations
in order to extract the weak charge from the experimental results. In the case of Cs, the 
atomic structure uncertainty is currently about 0.5\%. To go beyond this in heavy atoms 
probably requires measuring APV for a series of isotopes
of an atom and forming ratios
of the results. These ratios determine
$Q_W$ and are less sensitive to calculational uncertainties. However, 
this procedure is subject to other uncertainties
arising from variations in neutron distributions along the chain,
which will limit the ultimate precision\cite{erl03:prd}.

Apart from the potential of a precision measurement of $\sin^2 \theta_W$ and sensitivity
to new physics, experiments in hydrogen and deuterium provide the opportunity to determine all
four of the electron-nucleon, weak-neutral-current 
coupling constants: $C_{1p}$, $C_{2p}$, $C_{1n}$, and
$C_{2n}$ \cite{fei74:prd,fei74a:prd}. Here $C_{1p}$ and $C_{1n}$ are the nuclear
spin independent couplings for the proton and neutron.
(Note that $Q_W(p) \equiv 2C_{1p}$  \cite{bou74:pl,fei74a:prd}.)
$C_{2p}$ and $C_{2n}$ are the nuclear spin
dependent couplings. These four parameters are important
because they are basic properties of the neutron and proton. 
The constants for the neutron
can be obtained by comparing results from hydrogen and deuterium experiments. For
deuterium we use the notation \cite{note1:rwd}:
\begin{equation}
\eqalign{
C_{1D}&=C_{1p}+C_{1n}\cr
C_{2D}&=C_{2p}+C_{2n}.\cr
}
\label{deutdef}
\end{equation}
In terms of the quark coupling constants $C_{1u}$, $C_{1d}$, $C_{2u}$
and $C_{2d}$ 
(Ref.~\cite{yao06:jpg}),
the proton and deuteron couplings have the
values\cite{com80:arn,fay99:pva,mar83:prd,note4:rwd}:
\begin{equation}
\eqalign{C_{1p}&=-(2 C_{1u}+C_{1d}) \cr
C_{1D}&=-3(C_{1u}+C_{1d}) \cr
C_{2p}&=-0.935 C_{2u}+0.360 C_{2d} \cr
C_{2D}&=-0.45(1.1 C_{2u}+0.9 C_{2d}).
}
\label{quarks}
\end{equation}
The
constants $C_{1p}$ and $C_{1D}$ can be determined from
APV experiments with little strong 
interaction uncertainty. By contrast,
$C_{2p}$ and $C_{2D}$ 
involve considerable strong interaction uncertainty, as has
been discussed
by a number of authors\cite{mar83:prd,cah77:pl,col78:prd}.

In Table~\ref{limits} we give experimental and theoretical values
for these couplings. The experimental numbers in column 2 are 
based on a ``model-independent fit''
to all neutral current parameters which are presented in Table 10.8 of 
Ref.~\cite{yao06:jpg}. The fit allowed for
an arbitrary electroweak gauge theory. 
The final column of Table~\ref{limits} 
gives the SM predictions based on
Eq.~\ref{quarks}, Table 10.3 of Ref.~\cite{yao06:jpg}, and 
$\sin^2 \hat{\theta}_W
=0.23122$. 
Theoretical corrections from the
anapole \cite{khr00:npa} and strange quark \cite{cam89:plb} 
contributions are not included. 
All of the couplings except $C_{1D}$ are suppressed in the 
Standard Model.
For $C_{1p}$ and $C_{2p}$ the suppression arises because 
they are proportional
to $1-4\sin^2\theta_W$ in lowest order.
The coupling $C_{2D}$ vanishes exactly in
lowest order and the small residual value
is entirely due to higher order corrections, so this parameter
is quite sensitive to these corrections.
Although $C_{1D}$ has been
determined (indirectly) to about 3\%, the uncertainty in $C_{1p}$ is 
comparable to the size of the SM prediction, and 
both $C_{2p}$ and $C_{2D}$ are undetermined. 

An attractive goal for a future APV
experiment would be a 0.3\% measurement of the nuclear
spin independent coupling for the deuteron $C_{1D}$.
This would
determine $\sin^2\hat{\theta}_W$(0) with a combined error
 (theoretical + experimental) of
0.3\% which is a factor of three improvement
on the limit provided by the
heavy atom APV experiments. This is
comparable to the limit on
 $\sin^2\hat{\theta}_W$(0)
expected from
the Qweak experiment, but
would provide complementary
information on the isospin dependence
of the e-nucleon electroweak couplings \cite{you07:arx}.

A measurement of the nuclear spin independent
coupling in hydrogen at the
same level of absolute precision would provide
a 4\% measurement of $C_{1p}$, which together
with a measurement of $C_{1D}$ would provide a
constraint on the isospin dependence of the 
nuclear-spin-independent
weak-neutral-current interaction with
one experimental setup.
For $C_{2p}$ and $C_{2D}$, measurements
with absolute precisions similar to 
a 0.3\% measurement of $C_{1D}$ could
provide the first determinations of these quantities with an 
error of 3\% for $C_{2p}$
and 20\% for
$C_{2D}$. Again, the isospin dependence of 
the nuclear spin dependent couplings could
be constrained by comparing the two measurements.

\begin{table}
\caption{\label{limits}Parity violating weak neutral current couplings for H and D.}
\lineup
\begin{tabular*}{\textwidth}{@{}l*{15}{@{\extracolsep{0pt plus12pt}}l}}
\br
Coupling&From model-independent fit\cite{yao06:jpg}&Standard Model\\
\mr
$C_{1p}$&0.082$ \pm$0.034&0.0355\\
$C_{1D}$&\-0.441$ \pm $0.012&\-0.4587\\
$C_{2p}$&\-0.16\0$ \pm $0.26\0&0.043\0\\
$C_{2D}$&\-0.32\0$ \pm $0.4\0\0&0.007\0\\
\br
\end{tabular*}
\end{table}

\section{Parity mixing in hydrogen and deuterium\label{mixing}}

In planning parity experiments in
hydrogen, it is natural to take advantage of 
the metastable 2$^2$S$_{1/2}$ state. Metastable hydrogen atoms
live long enough ($\sim$ 122 ms) to survive the length of a
typical laboratory atomic beam apparatus, and because of 
the near degeneracy of the 2$^2$S$_{1/2}$ and 2$^2$P$_{1/2}$
states, level crossings occur at 
relatively modest magnetic fields ($\sim$ kG), 
and this 
provides a means for separately determining the weak
interaction coupling constants, since different 
combinations of constants are resonantly
enhanced at the various level crossings. 
In order to determine all four coupling
constants in 
Table~\ref{limits}, it is necessary
to perform experiments 
at several magnetic field settings in both H and D,
and combine the results to isolate the separate
constants.
In this section, we review
the parity mixing matrix elements
at the various crossings in 
H and D and present updated SM predictions
for their strengths.

\begin{figure}
\includegraphics[width=3.4in]{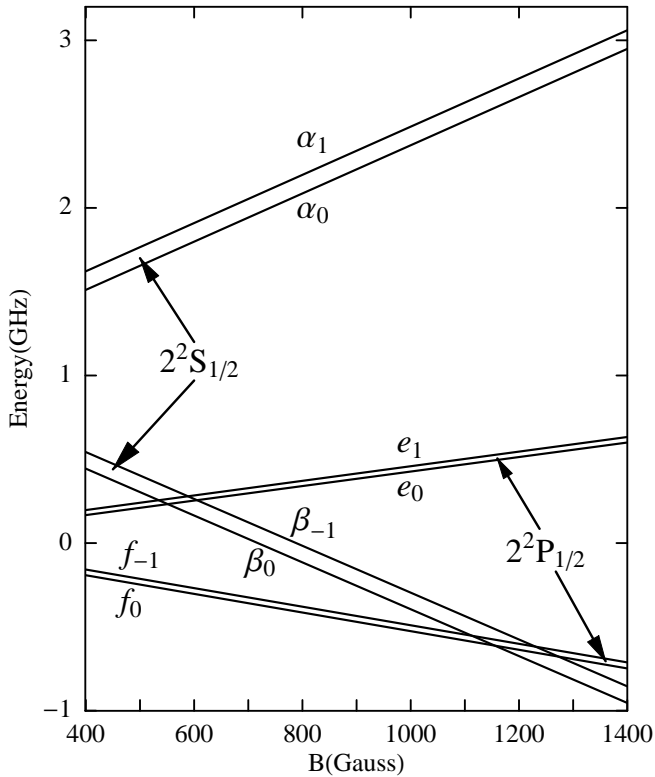}
\caption{Zeeman diagram showing the 2$^2$S$_{1/2}$
and 2$^2$P$_{1/2}$ levels in hydrogen in a static
magnetic field\cite{lev82:dis}. The states are labeled using
Lamb's nomenclature
\cite{lam52:pr}. Subscripts give the values
of the good quantum number m$_{\rm F}$.
}
\label{fig:zeeman}
\end{figure}

Fig.~\ref{fig:zeeman} shows the relevant 
energy levels of hydrogen in a static magnetic field
of 400-1400 G.
In this region, the Zeeman
energy from the coupling to the magnetic field is
much less than the fine structure splitting so
the total electronic angular momentum J is still a
good quantum number. On the other hand, at fields above
$\sim$ 200 G, the electron and nuclear spins are 
decoupled so the natural basis is
$|{\rm J} {\rm m_J I m_I} \rangle$, where I is the nuclear spin.
In Fig~\ref{fig:zeeman} the 2$^2$S$_{1/2}$ levels
are labeled $\alpha$ (m$_{\rm J}$=+1/2) and $\beta$
(m$_{\rm J}$=$-$1/2), while the 2$^2$P$_{1/2}$ states
are labeled e (m$_{\rm J}$=+1/2) and
f  (m$_{\rm J}$=$-$1/2)
\cite{lam50:pr}.
The subscripts are 
m$_{\rm F}$ = m$_{\rm J}$ + m$_{\rm I}$.

\begin{figure}
\includegraphics[width=3.4in]{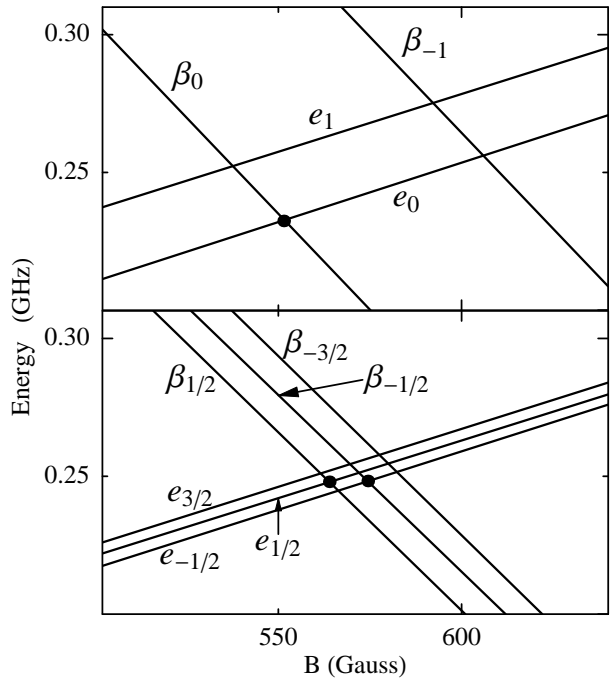}
\caption{Parity mixings resonant at the $\beta e$ crossings in hydrogen
(upper part) and deuterium (lower part).}
\label{fig:be}
\end{figure}

Fig.~\ref{fig:be} shows blowups of the   
$\beta$e level crossings in hydrogen and deuterium. 
The selection
rule for H$_{\rm PNC}$ is
$\Delta {\rm m}_{\rm F}=0$, so only the levels $\beta_0$
and e$_0$ in hydrogen are mixed.
The value of the matrix element is given
in Table~\ref{crossings}, which indicates
that a measurement at the $\beta$e crossings
in hydrogen is
sensitive only to the nuclear spin
dependent coupling C$_{2p}$.
In deuterium, 
parity mixing 
is resonant at two $\beta$e crossings 
which, as indicated in Table~\ref{crossings}, 
are sensitive only to $C_{2D}$.

\begin{figure}
\includegraphics[width=3.4in]{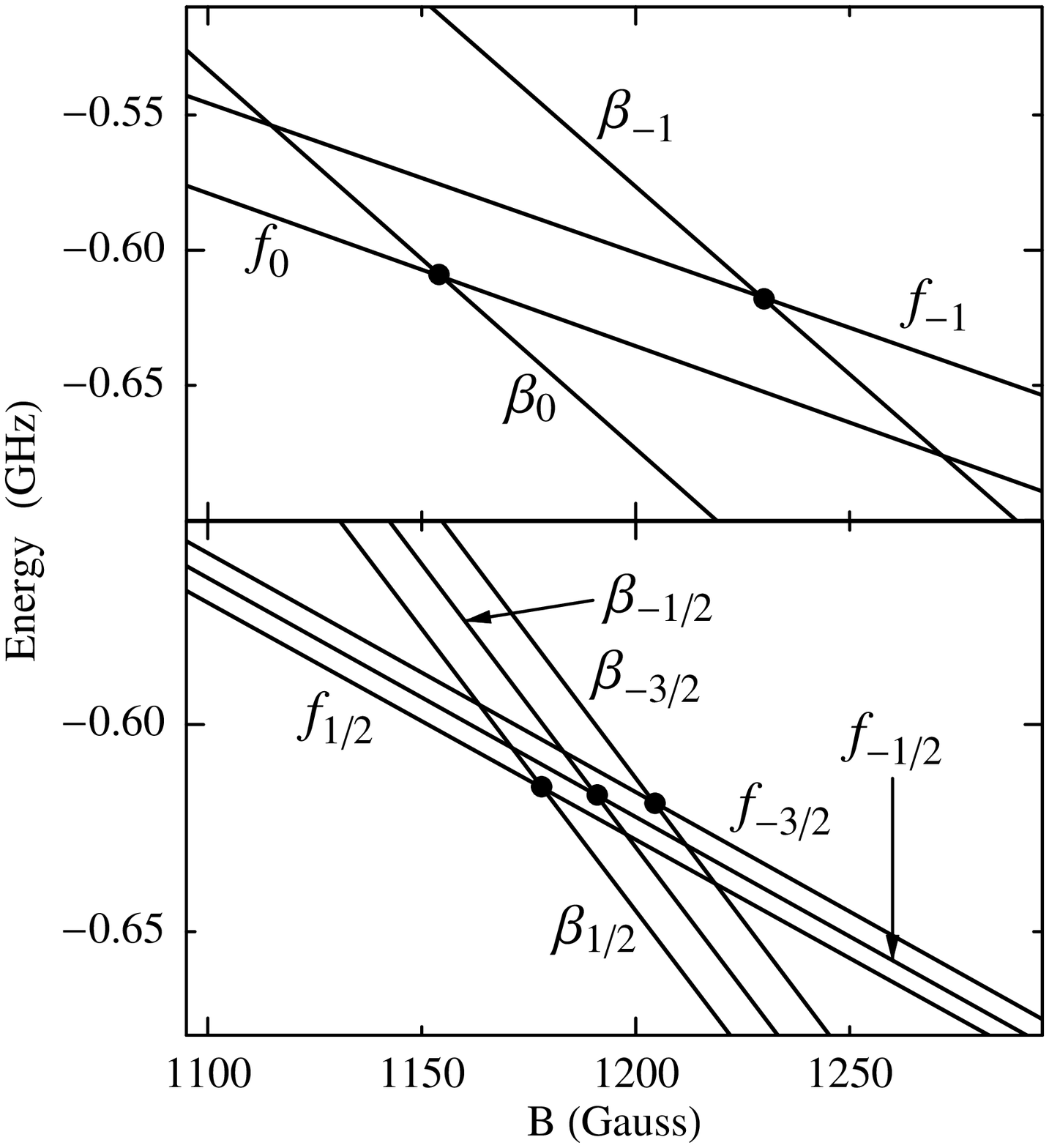}
\caption{Parity mixings resonant at the $\beta f$ crossings in hydrogen (upper part)
and deuterium (lower part).}
\label{fig:bf}
\end{figure}

Fig.~\ref{fig:bf} shows 
the $\beta$f level crossings in
hydrogen and deuterium. Parity mixing is resonant at 
two crossings in hydrogen. Table~\ref{crossings} indicates
the matrix elements depend on nearly orthogonal
combinations of $C_{1p}$ and $C_{2p}$. 
So parity experiments designed to 
use the $\beta$f crossings in hydrogen provide a means
to measure both of these constants. For the
$\beta$f crossings in deuterium,
there are three crossings
of levels mixed by the weak interaction. From 
Table~\ref{crossings} it is seen that two
of these depend on orthogonal combinations
of $C_{1D}$ and $C_{2D}$ while the 
crossing of $f_{-1/2}$ and $\beta_{-1/2}$
is, to good approximation,
sensitive only to $C_{1D}$.

\begin{table}
\caption{\label{crossings}Matrix elements for parity-violating,
weak-neutral-current couplings for
hydrogen and deuterium at the level crossings. In units of
$\rmi \bar{\mathcal{V}}_{\rm{w}} \approx 2\pi$ $\times$ $0.013$ $ s^{-1}$.
$C_{1D} \equiv C_{1p}+C_{1n}$ and $C_{2D} \equiv  C_{2p} + C_{2n}$.}  
\lineup
\begin{tabular*}{\textwidth}{@{}l*{15}{@{\extracolsep{0pt plus12pt}}l}}
\br
Matrix Element&Value&Standard Model\\
\mr
Hydrogen \\
\ms
$\langle e_{0} |V|\beta_{0} \rangle$&$\-2 C_{2p}$&\-0.086\\
$\langle f_{0} |V|\beta_{0} \rangle$& $C_{1p}+1.1C_{2p}$&0.083\\
$\langle f_{-1} |V|\beta_{-1} \rangle$&$C_{1p}-C_{2p}$&\-0.007\\
\bs
Deuterium \\
\ms
$\langle e_{1/2} |V|\beta_{1/2} \rangle$&$\-\sqrt{2} C_{2D}$&\-0.010\\
$\langle e_{-1/2} |V|\beta_{-1/2} \rangle$&$\-\sqrt{2} C_{2D}$&\-0.010\\
$\langle f_{1/2} |V|\beta_{1/2} \rangle$&$C_{1D}+C_{2D}$&\-0.452\\
$\langle f_{-1/2} |V|\beta_{-1/2} \rangle$&$C_{1D}$&\-0.459\\
$\langle f_{-3/2} |V|\beta_{-3/2} \rangle$&$C_{1D}-C_{2D}$&\-0.466\\
\br
\end{tabular*}
\end{table}

\section{Current limits on parity violation in
hydrogen\label{currentlimits}}

There have been numerous suggestions for APV experiments in
hydrogen. One possibility is to observe PNC asymmetries in the
single photon decay 2S$_{1/2}$ $\rightarrow$ 1S$_{1/2}$ + $\gamma$
of a metastable beam in vacuum\cite{fei74:prd}, or 
in a beam perturbed by 
applied electric and magnetic fields\cite{azi75:jet,dun81:pra}.
These ideas have not led to practical experiments in hydrogen 
although they hold promise for experiments in heavy one-electron
ions\cite{dun81:pra}. For hydrogen, the most 
promising approach is to drive
optical or microwave transitions 2S$_{1/2}$ $\rightarrow$ 
nS$_{1/2}$ and look for a pseudoscalar dependence on the
transition rate\cite{lew75:pl,hin77:pl,dun78:pra,dun78:dis,
ade81:nim}. Microwave (or RF\cite{hin88:sah}) transitions
2s $\rightarrow$ 2s' within n=2 are particularly attractive 
because
Doppler broadening is negligible, and linewidths can be small.
Also, relatively high
power is readily available and the interaction volume
can be large.
The possibility for narrow linewidths is particularly
significant since it allows complete separation of the various hyperfine
transitions, and, as discussed in Section \ref{sensitivity}, 
a narrow line is required to optimize the sensitivity
to parity violation.

Starting in the late 1970s, hydrogen parity experiments
were performed at the University
of Michigan \cite{lew75:pl,dun78:pra,lev82:prl,lev84:pra,feh93:dis}, 
The University of Washington \cite{ade81:nim,chu83:dis},
and Yale University \cite{hin77:pl,hin88:sah}. 
The experiments were motivated in part by the near
degeneracy of
the 2$^2{\rm S}_{1/2}$ and 2$^2{\rm P}_{1/2}$ levels in hydrogen
which enhances the parity mixing and
provides an opportunity
for detection of 
PNC effects even though hydrogen does
not benefit from the Z$^3$ enhancement 
as a function
of nuclear charge which motivates 
parity experiments in heavy atoms\cite{khr91:gab}.

All of the hydrogen APV experiments 
were based on fast ($\sim$ 500 eV) beams of 
metastable hydrogen atoms and aimed to
measure $C_{2p}$. Typically,
the beam was prepared in one of the $\alpha$
hyperfine states and a transition
$\alpha \rightarrow \beta$
to one of the $\beta$ hyperfine states
was driven.
The transition was 
observed by selective quenching of the 
resulting $\beta$
state atoms and detecting the Lyman-$\alpha$
radiation emitted.

The transitions were driven by
a parity conserving amplitude A$_{\rm PC}$
and a parity nonconserving amplitude
A$_{\rm PNC}$. The
parity nonconserving amplitude proceeds $via$ an
intermediate 2$P_{1/2}$
state:
\begin{equation}
\raisebox{.01pt}{\normalsize{$\alpha$}}
\;\; \longrightarrow \hspace{-1.8em}
\raisebox{9pt}{\bf \large{$\epsilon$}}
\raisebox{14pt}{\small{$\omega$}}
\:\:\:\; 2P_{1/2} \:
\longrightarrow \hspace{-2.2em}\:
\raisebox{9pt}{V}\raisebox{8pt}{\small{w}} \;\;\;
\raisebox{0.01pt}{\normalsize{$\beta$}},
\label{apnc}
\end{equation}
where the first step is an $E1$
transition driven by microwave electric
field $\epsilon^{\omega}$, 
and the second step is the 2$^2{\rm S}_{1/2}-
2^2{\rm P}_{1/2}$ coupling by the
weak interaction. This is also 
called a ``weak induced amplitude''.

Two types of parity conserving amplitudes A$_{\rm PC}$
were used. One of these was an M1 amplitude\cite{hin77:pl}:
\begin{equation}
\raisebox{.01pt}{\normalsize{$\alpha$}}
\;\; \longrightarrow \hspace{-1.7em}
\raisebox{9pt}{\bf \normalsize{$b$}}
\raisebox{15pt}{\small{$\omega$}}\;\;\;
\raisebox{0.01pt}{\normalsize{$\beta$}},
\label{apcm1}
\end{equation}
where $b^{\omega}$ is the microwave 
magnetic field. More commonly
a ``Stark induced amplitude'' was
used:
\cite{dun78:pra,ade81:nim}:
\begin{equation}
\raisebox{.01pt}{\normalsize{$\alpha$}}
\;\; \longrightarrow \hspace{-1.8em}
\raisebox{9pt}{\bf \large{$\epsilon$}}
\raisebox{14pt}{\small{$\omega$}}
\:\:\:\; 2^2{\rm P}_{1/2} \:
\longrightarrow \hspace{-2.2em}\:
\raisebox{9pt}{$E$}\raisebox{14pt}{\small{s}} \;\;\;
\raisebox{0.01pt}{\normalsize{$\beta$}},
\label{apce1}
\end{equation}
here, the first step is a 
microwave $E1$ coupling to an intermediate
2$^2{\rm P}_{1/2}$ state driven by $\epsilon^{\omega}$,
and the second step is a 2$^2{\rm S}_{1/2}-
2^2{\rm P}_{1/2}$ coupling by a static
electric field $E^s$.

The experiments searched for a
parity violating (pseudoscalar) interference term
between A$_{\rm PNC}$ and A$_{\rm PC}$ 
in the transition rate, that
changed sign as the 
``handedness'' of the interaction region 
was switched. The main differences
among the various schemes were in 
the configurations of fields in the interaction
regions, and the 
pseudoscalars of interest.
Particularly significant is that in the Michigan experiment,
A$_{\rm PC}$ and A$_{\rm PNC}$ were
driven simultaneously in a single microwave
cavity, while in the Yale and U. Washington
experiments, separate oscillating
field regions were used for A$_{\rm PNC}$ and A$_{\rm PC}$. 

These experiments were not able to observe parity
violation (for a short review see \cite{dun99:pva}). 
The best limit was obtained by
Fehrenbach \cite{feh93:dis} who reported:
\begin{equation}
C_{2p}=1.5 \pm 1.5 \pm 22.
\label{fehrenbach}
\end{equation}
Here, the first error is statistical and the second is systematic.
From Table~\ref{limits}, we see that
the value expected based on the 
Standard model is C$_{2p}$= 0.036. So the 
statistical error is
a factor of 40 larger than the expected
effect while the systematic
error is 600
times larger than the expected
effect. It is clear
that more than an incremental improvement
is necessary if a 
useful measurement is to be obtained. 
Some ideas of how this might be accomplished are
given in sections \ref{sensitivity} and 
\ref{systematics}.

\section{Sensitivity of a generic deuterium APV 
experiment \label{sensitivity}}

The original hydrogen parity experiments could not have observed PNC.
The problems were a lack of sufficient sensitivity 
and large systematic errors. 
In the next two sections we discuss
a particular idea 
for an APV experiment in deuterium 
aimed at providing a 0.3\% measurement
of C$_{1D}$. 
The conclusions of this paper do not
rest on this particular scheme, as
there are many other possibilities
which would have roughly the same
sensitivity. Rather, we chose this 
example after considering
some general guidelines
for a future H/D parity experiment. First,
in comparison to the original H APV 
experiments, the
signal-to-noise-ratio S/N  
has to be improved by more than two orders of
magnitude. Then, to improve handling of 
systematic effects, it is important 
to have precise 
control
over the alignment of the apparatus and 
all applied fields.
To make this less daunting, the apparatus 
should be simple -- 
a few homogeneous
applied fields is preferred over
more complicated configurations. 
Perturbations of the beam due to 
surfaces (apertures, walls, etc.) should be minimized,
and all surfaces should be kept as far from the
beam as possible.
The apparatus
should be ``clean'' $i.e.$ surfaces should be carefully
prepared to minimize stray electric fields, and ultra-high
vacuum techniques should be used.
The beam density should be as low as possible
consistent with the needed intensity and there should be no
charged particles, or
field
configurations which tend to trap charged particles.

One can increase S/N
by developing a more intense 
metastable beam. In addition, there are advantages
in both sensitivity, and in reducing systematic
errors to using a slow 
beam\cite{ade81:nim,lev84:pra}.
In \ref{secopticalpumping} we discuss 
the requirements for producing a slow (77K) metastable
beam with about
5 $\times$ 10$^{14}$ D(2s)s$^{-1}$ 
per hyperfine level.
An important consideration 
is that there is a limit on the density
of metastable atoms in the beam
due to quenching by beam-beam
collisions. This is discussed in
\ref{secquenching}. 
\begin{figure}
\includegraphics[width=2.4in]{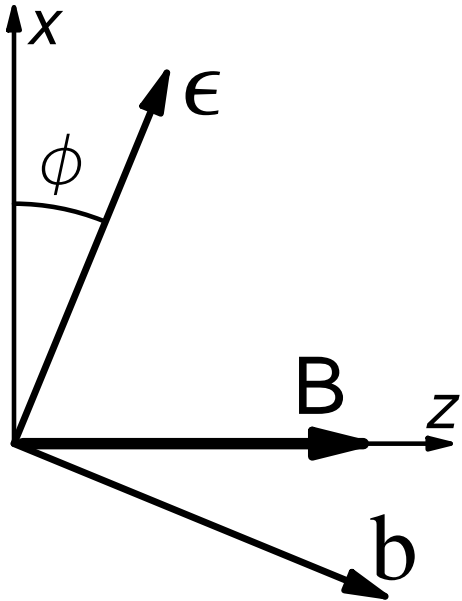}
\caption{Fields in the interaction region. 
The metastable D beam is travelling along the z axis.
 The microwave electric $\bm{\epsilon}$ and magnetic
 ${\bf b}$ fields lie in the x-z plane. ${\bf B}$ is
 a static magnetic field.}
\label{fig:polphi}
\end{figure}

A particularly simple configuration for the 
interaction region 
would involve driving an M1 transition
\cite{hin77:pl} with
a plane polarized {\em microwave} {\em beam} 
incident at right
angles to the metastable beam. 
This would eliminate the need for
an additional applied electric
field to drive
A$_{\rm PC}$.
Also, eliminating
the microwave cavity 
(or cavities), has some  advantages.
The standing wave pattern in a cavity leads
to spatial variations in the fields which
can complicate the transition amplitudes 
\cite{rob82:jap} and
make analyzing systematic
errors more difficult.
In addition, apertures 
at the entrance or exit to the cavity
perturb both the 
atoms and the cavity modes.

A practical way to realize 
this is 
to use a microwave-open-ring-resonator
\cite{sch67:mtt,men73:pra,bal81:rsi}
in the interaction region.
Complete control of the polarization
in such a resonator is possible using
quasi-optical techniques\cite{dun86:pra}.
Such a resonator
allows the propagation
direction of the traveling waves to
be switched, providing the
ability to reverse the
microwave propagation vector ${\bf k}$.

\begin{figure}
\includegraphics[width=3.4in]{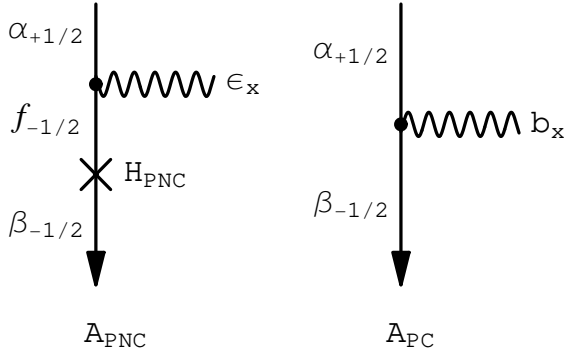}
\caption{Parity conserving ${\rm A_{PC}}$ and nonconserving 
${\rm A_{PNC}}$ amplitudes.}
\label{fig:ampli}
\end{figure}

As an example, we consider a
measurement of $C_{1D}$ at the crossing of the levels
$f_{-1/2}$ and $\beta_{-1/2}$, 
utilizing the configuration of fields
shown in 
Fig.~\ref{fig:polphi}. 
The D(2s) beam is directed 
along the
z axis, collinear with a static
magnetic field ${\bf B}$. 
The beam of microwaves
propagates into the page
$({\bf \hat{k}} = -{\bf \hat{y}})$
and the
frequency is set to drive the
M1 
transition
$\alpha_{+1/2} 
\rightarrow \beta_{-1/2}$.
The microwave electric polarization 
vector $\bm{\epsilon}$ lies in the x-z plane,
and is orthogonal to the microwave 
magnetic field ${\bf b}$. 
The vector $\bm{\epsilon}$ is a combination of
fields $\bm{\epsilon}_x$ and 
$\bm{\epsilon}_z$ \cite{dun86:pra}.
These fields are derived from the same source but
have a phase difference of $\theta$ and separately
controlled amplitudes.
The resulting field is given by:
\begin{equation}
\bm{\epsilon}(t)=\bm{\epsilon}_x\cos(\nu t)+\bm{\epsilon}_z\cos(\nu t+\theta).
\label{eqb1}
\end{equation}
Here, we will set $\theta = 0$ or $\pi$, in which case, the microwave 
polarization is linear. The polarization 
angle $\phi$ relative to the $x$ axis is determined by the relative strength of 
$\bm{\epsilon}_x$ and $\bm{\epsilon}_z$. When $\theta$
is changed from 0 to $\pi$, $\phi$ goes to -$\phi$. Other values of 
$\theta$ give circular or elliptical polarization. The amplitudes in the 
experiment are shown schematically in Fig~\ref{fig:ampli}. The parity conserving 
part ${\rm A_{PC}}$ is an $M1$ amplitude driven by
the component of the microwave magnetic 
field along the $x$ axis (b$_x$).  

The parity nonconserving amplitude ${\rm A_{PNC}}$ 
proceeds $via$ the intermediate
level $f_{-1/2}$ and involves the weak interaction H$_{\rm PNC}$
and an $E1$ coupling via the 
microwave electric field component along the $x$ axis ($\epsilon_x$).
The analysis of this experiment requires consideration of the equations
of motion for the three levels $\alpha_{+1/2}$, $\beta_{-1/2}$, and $f_{-1/2}$ but,
to sufficient approximation for the present considerations,
can be reduced to an effective two-level problem by
introducing the parameter $\gamma_{\alpha_{+1/2}}$ which is the
decay rate of the $\alpha_{+1/2}$ level due to the quenching
$\alpha_{+1/2} \rightarrow f_{-1/2}$ by the microwaves\cite{lev84:pra}. The
amplitudes $a$ for the $\alpha_{+1/2}$ state and $b$ for the $\beta_{-1/2}$ 
state obey
the equations:
\begin{equation}
\eqalign{\rmi \dot{a}=\Lambda^{\dagger} \exp[-
\rmi(\omega_{\alpha_{+1/2}\beta_{-1/2}}-\nu)t]
b-\case{1}{2} \rmi \gamma_{\alpha_{+1/2}} a \cr
\rmi \dot{b}=\Lambda \exp[+\rmi(\omega_{\alpha_{+1/2}\beta_{-1/2}}-\nu)t] 
a}
\label{eqb2}
\end{equation}
where\cite{note7:rwd}:
\begin{equation}
2\hbar \Lambda = -\mu_0 \rm{b}_{\it x}  - 
\rmi \bar{\mathcal{V}}_w
C_{1D}
{\it  d}_2 \epsilon_{\it x} \Delta_{\beta_{-1/2} f_{-1/2}}.
\label{eqb3}
\end{equation} 
The parameter $\bar{\mathcal{V}}_{\rm{w}} =2\pi (0.013)s^{-1}$
gives the magnitude of the weak interaction matrix element,
and $d_2$=$\sqrt{3}\it{ea}_0$.
The parameter
$\Delta_{\beta_{-1/2} f_{-1/2}}$
is defined by:
\begin{equation}
\Delta_{ij}= \frac{1}
{\omega_{ij}+
\rmi\gamma_{2 \it p}/2}.
\label{eqb3a}
\end{equation}
Here, $\omega_{ij}\equiv\omega_i - \omega_j$,
and $\gamma_{2p}=2\pi\times10^8s^{-1}$ is the natural decay
rate of the $2p$ state.
The beam is prepared in the $\alpha_{+1/2}$ state
so the initial conditions are:
\begin{equation}
\eqalign{a(0)=1 \cr
b(0)=0.}
\label{eqb4}
\end{equation}
For simplicity, we treat the case of exact resonance
$\omega_{\alpha_{+1/2}\beta_{-1/2}} -
\nu = 0$, and also choose the maximum interference between the two
terms in Eq.~\ref{eqb3} which requires $\omega_{\beta_{-1/2}f_{-1/2}}$=0,
i.e. that the magnetic field be tuned
to the center of the level crossing. 
If the atom spends a time $\tau$ in the interaction
region, the
amplitude to be in the state $\beta_{-1/2}$ after 
this time is given by:
\begin{equation}
b(\tau)=-\rmi \Lambda \frac{1 - \exp(-\gamma_{\alpha_{+1/2}}\tau/2)}
{\gamma_{\alpha_{+1/2}}/2}.
\label{eqb5}
\end{equation}
The two terms in $\Lambda$ 
(Eq.~\ref{eqb3}) determine the two amplitudes of interest.
${\rm A_{PC}}$ is proportional to the
first term
while ${\rm A_{PNC}}$ is proportional to the second.
Since the electric $\bm\epsilon$ and magnetic
$\bf b$ fields in the plane wave are orthogonal,
$\rm b_{\it x}=b\sin \phi$ and $\epsilon_x=\epsilon \cos\phi$, so the
sign of the interference term changes as
$\phi \rightarrow - \phi$. 
The interference term is proportional to
the pseudoscalar
$(\bm{\epsilon} \cdot \bf{B})
(\bm{\epsilon} \times \bf{k}) \cdot \bf{B}$,
so it also changes sign
if the microwave propagation vector is reversed,
i.e. $\bf{k} \rightarrow -\bf{k}$; but it is
even under a reversal of $\bf{B}$.
This pseudoscalar can appear in a T-even theory
due to the presence of damping of the 2p state
\cite{dun78:dis,lev82a:prl,dun97:prl}.

The parameter $\gamma_{\alpha_{+1/2}}$ is 
determined in a separate calculation. If the 
magnetic field is set at the level crossing 
and the microwave frequency is set at the center
of the $\alpha_{+1/2} \rightarrow \beta_{-1/2}$ resonance we have\cite{lam52:pr}:
\begin{equation}
\gamma_{\alpha_{+1/2}}=\gamma_{2p}\left(\frac{d_2 \epsilon_x}
{\hbar \gamma_{2p}}\right)^2
\label{eqb6}
\end{equation}

Two quantities are useful in assessing the sensitivity of this type of APV  
experiment. The first is the fraction of the rate that is due to the PNC 
interference term, which
we will call the asymmetry $\mathcal{A}$:
\begin{eqnarray}
\mathcal{A}&=4\frac{\bar{\mathcal{V}}_{\rm w}}{\gamma_{2p}}\frac{d_2 \epsilon_x}{\mu_0 {\rm b}_x}
C_{1D} \label{eqb7}\\
&\approx \frac{-1.1 \times 10^{-7}}{\tan\phi},\label{eqb8}
\end{eqnarray}
where we used $C_{1D}$ = $-$0.459, and
$b_x/\epsilon_x= \tan \phi$. The 
asymmetry can be set experimentally by 
choosing the angle $\phi$. Here we just
assume the value $\tan \phi=0.1$ which
gives an asymmetry of $-1.1 \times 10^{-6}$.

\begin{table}
\caption{\label{definitions}Parameters determining the
optimized sensitivity 
(Eq.~\ref{eqb11})}
\lineup
%\begin{tabular*}{\textwidth}
%{ @{}l*{15}{ @{ \extracolsep{0pt plus12pt} }l } }
\begin{tabular*}{\textwidth}{cll}
\br
Symbol&Meaning& Assumed for Figs.~\ref{fig:cdsens} and \ref{fig:cpsens}\\
\mr
$\eta$&Total detection efficiency&0.5\\
$J$&Metastable flux/hyperfine component&
$-$\\
$T$&Counting time&5 days\\
$\tau$&Interaction time&$-$\\
$\rmi \bar{\mathcal{V}}_{\rm w}$ &H PNC
coupling strength (n=2)& $2\pi$ $\times$ $0.013$ $s^{-1}$\\
$\gamma_{2p}$ & 2p decay rate &$2\pi$ 
$ \times$ $10^{8}$ $s^{-1}$\\
\br
\end{tabular*}
\end{table}

A more important quantity for assessing the viability
of H/D APV experiments
is the counting time required under ideal conditions, 
$i.e.$ assuming one can achieve the shot-noise limit, and
that the transition rate is well above background.
 The signal count $S$ 
is proportional to $(A_{\rm PNC}^*A_{\rm PC}+\rm{c.c.})$ times
the factor $\eta J T$, where $\eta$ is the
detection efficiency for the 
$\beta_{-1/2}$ state, $J$ is the 
metastable beam intensity per
hyperfine component (atoms/s), 
and $T$ is the total counting
time. The noise count $N$ is
proportional to the square root of $\eta J T | {\rm A_{PC}} |^2$. Using
Eqs.~\ref{eqb3}, \ref{eqb5} and \ref{eqb6} we find:
\begin{equation}
\frac{S}{N}=4|\bar{\mathcal{V}}_{\rm w} C_{1D}|
\frac{1-\exp(-\gamma_{\alpha_{+1/2}}\tau/2)}
{(\gamma_{\alpha_{+1/2}}\gamma_{2p})^{1/2}}
(\eta J T)^{1/2}.
\label{eqb9}
\end{equation}
Note that Eqs.~\ref{eqb9} does not depend on $\phi$ and, in
the approximation we use here, the counting time is independent of the
asymmetry chosen for the experiment. 
In fact, as pointed out by Hinds\cite{hin80:prl},
if the phase between A$_{\rm PC}$ and
A$_{\rm PNC}$ is chosen to maximize
the interference term, the signal-to-noise
is independent of
A$_{\rm PC}$.

Equation~\ref{eqb9} has a maximum with respect $\gamma_{\alpha_{+1/2}}$
given by the condition:
\cite{hin80:prl}:
\begin{equation}
\gamma_{\alpha_{+1/2}}=2.5/\tau,
\label{eqb10}
\end{equation}
which, for a given $\tau$, can be satisfied 
by adjusting the microwave power. Since the required
power is readily obtained for the experiments considered
here, we use Eq.~\ref{eqb10} to eliminate $\gamma_{\alpha_{+1/2}}$ from
Eq.~\ref{eqb9} with the result:
\begin{equation}
\left(\frac{S}{N}\right)_{opt}=1.8 |\bar{\mathcal{V}_{\rm w} }C_{1D}|
(\eta J T \tau/\gamma_{2p})^{1/2}.
\label{eqb11}
\end{equation}
The definitions of the parameters in this equation are
summarized in Table~\ref{definitions}.
Eq.~\ref{eqb11} 
indicates that,
apart from maximizing the beam intensity and
detection efficiency, making $\tau$ as large
as possible is the key requirement for 
maximizing S/N.
This means an experiment based on
a slow beam and a long interaction region. A hidden
assumption here
is that the linewidth of the $\alpha_{+1/2}\rightarrow
\beta_{-1/2}$ transition is of order the width
of the $\alpha_{+1/2}$ state (Eq.~\ref{eqb6}).
In particular, the 
magnetic field must be constant and homogeneous
at the level:
\begin{equation}
\frac{\Delta B}{B}<\frac{\gamma_{\alpha_{+1/2}}}
{\omega_{\alpha_{+1/2}\beta_{-1/2}}}\sim\frac{10^{-10}s}{\tau}.
\label{eqb12}
\end{equation}

\begin{figure}
\includegraphics[width=3.4in]{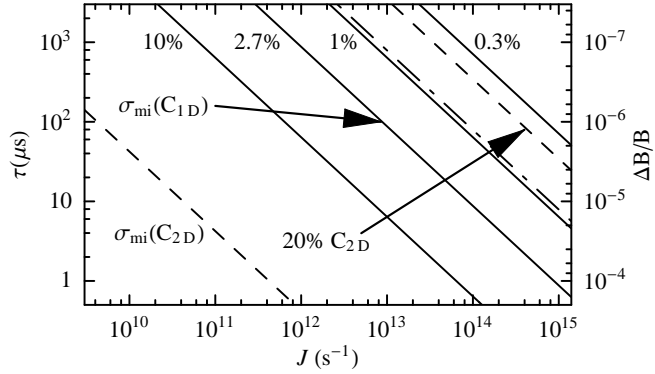}
\caption{Deuterium APV sensitivity.
Solid lines indicate projected minimum experimental
uncertainties in $C_{1D}$ 
(in percent of the SM value)
for a generic APV experiment in D 
based on Eq.~\ref{eqb11}. 
These are plotted as a function
of the time an atom spends in 
the interaction region $\tau$, 
and the beam intensity per hyperfine component
$J$. The values assumed for the 
other parameters in Eq.~\ref{eqb11}
are summarized in Table~\ref{definitions}.
The dashed lines are the 
projected uncertainties for a $C_{2D}$ 
measurement at the $\beta$e crossing. The dot-dash line
represents the uncertainty needed in 
a measurement of $C_{1D}$ 
to determine $\sin^2 \theta_w$ with a precision
equal to the current APV limit of 0.9\%. 
The lines marked $\sigma_{\rm mi}(C_{2D})$
and $\sigma_{\rm mi}(C_{1D})$ represent the 
uncertainties in these parameters
based on a model-independent-fit
to all neutral current data\cite{yao06:jpg}.
The axis on the right gives the magnetic field
homogeneity $\Delta B/B$ based on Eq.~\ref{eqb12}.
}
\label{fig:cdsens}
\end{figure}

\begin{figure}
\includegraphics[width=3.4in]{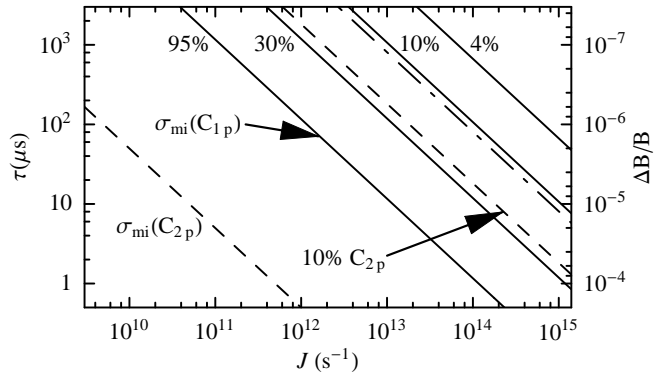}
\caption{Hydrogen APV sensitivity. 
Projected minimum experimental uncertainties
(as a percentage of the SM values)
in the hydrogen weak interaction coupling constants
based on Eq.~\ref{eqb11}, and plotted as
a function of $\tau$ and $J$.
(See caption to Fig.~\ref{fig:cdsens}.)
The solid lines refer to 
a measurement of $C_{1p}$ and the
dashed lines refer to measurement of $C_{2p}$
at the $\beta$e crossing. The dot-dashed
line corresponds to the uncertainty in $C_{1p}$ 
required to match
the current APV limit on $\sin^2 \theta_w$.
The lines marked $\sigma_{\rm mi}(C_{1p})$ and $\sigma_{\rm mi}(C_{2p})$
correspond to the uncertainties in these parameters
determined by
the model-independent-fit discussed
in Ref.~\cite{yao06:jpg}.
}
\label{fig:cpsens}
\end{figure}

Eq.~\ref{eqb11} also applies to measurements performed at the
other level crossings listed in Table~\ref{crossings}, providing
the combinations of coupling constants listed there
are substituted in place of $C_{1D}$. 
We can use the optimized signal-to-noise ratio to estimate 
the minimum limits that can be placed on the electron-nucleon, 
weak-neutral-current coupling constants from
experiments in H and D. To do this, we assume that
the counting time $T$ should not be
more than about 5 days; with the 
understanding that, in reality, considerably
more counting time would be required because of 
backgrounds and the need to study systematic
errors. Then, taking the overall detection 
efficiency to be $\eta=0.5$, we plot the percent errors in the
four constants as a function of the metastable beam intensity per
hyperfine component $J$, and the time of passage of an atom through the
interaction region $\tau$. We also
give the magnetic field
homogeneity required. The results for the deuterium couplings are
given in Fig.~\ref{fig:cdsens}, and those for hydrogen are given in 
Fig.~\ref{fig:cpsens}. For example,  Fig.~\ref{fig:cdsens} indicates
that to measure $C_{1D}$ to $\sim $ 0.3\% with 
an interaction time $\tau \sim 100 \mu s$, requires a beam intensity
of at least
$J \simeq$ 5 $\times$ 10$^{14}$s$^{-1}$ and $\Delta B/B \sim 10^{-6}$.
For longer interaction times
$\tau$, the requirement on the beam intensity can be reduced
but this
also leads to more stringent
specifications on the B-field homogeneity. 
The same experimental conditions would provide a measurement of
$C_{2D}$ to better than $20\%$ and (based on Fig.~\ref{fig:cpsens})
a $\sim 4\%$ measurement of $C_{1p}$.
``Interesting'' experiments can be done with smaller
beam fluxes as can be seen from the lines in Figs.~\ref{fig:cdsens}
and \ref{fig:cpsens} indicating the model-independent limits on the 
four coupling constants as well as the ``dot-dashed'' lines
which show the experimental
uncertainty required to match the current APV limit
on determination of $\sin ^2\theta_w$.

%Eq.~\ref{eqb11} can be used to give an estimate
%of the sensitivity of
%other H PNC experiments done at the $2S_{1/2}-2P_{1/2}$
%level crossings and in particular, it applies
%to the former experiments
%\cite{hin77:pl,ade81:nim,lev82:prl} done
%at the $\beta e$ crossing (with $C_{1p}+1.1C_{2p}\rightarrow
%-2C_{2p}$). These experiments used a fast beam with
%typical parameters $\tau=0.3\mu s$
%and $J\sim 10^{13}s^{-1}$ resulting in
%about a factor of $\sim$60 smaller signal to noise ratio, 
%so could not have observed H PNC
%in a practical length of time.

\section{Systematic Effects\label{systematics}}
The $2^2{\rm S}_{1/2}$ and $2^2{\rm P}_{1/2}$ levels in hydrogen 
can be mixed by unwanted electric fields and this
can mimic parity mixing. The closeness of the
levels enhances the PNC effects  
but it also increases the sensitivity of the experiments
to systematic effects due to unwanted electric fields. 
In this section, we will consider two of the most 
troublesome effects expected in the experiment
discussed in Section~\ref{sensitivity}.

A number of essential elements of the
treatment of systematic errors will not be
covered here, however. These include
the techniques for accurately aligning
the apparatus, and reversing 
the vectors in the interaction region
including: the polarization
$\phi$, the microwave propagation $\bf k$,
and the magnetic
field ${\bf B}$. We will assume that
the apparatus can be aligned to the
levels achieved
in the former
experiments $e.g.$ \cite{feh93:dis}.
However, in \ref{norm}  
a scheme for continuously monitoring the
microwave polarization is discussed, that
could make this reversal exact to
the level of statistical accuracy.
Another crucial issue
concerns effects
caused by variations in the intensity
or direction of the fields as the atoms
move through the apparatus\cite{rob82:jap}. 
These effects
will not be discussed here. They are addressed
in detail by $e.g.$
L\'evi and Williams\cite{lev84:pra}.

To understand the problem
of unwanted electric fields, we introduce 
electric fields $E_x^u$, $E_y^u$ and $E_z^u$ into our amplitudes.  
We also allow a misalignment of the 
microwave field which gives components $\epsilon_y^m$ 
and b$_y^m$ in addition to the desired fields.
The magnetic field defines
the $z$ axis and we are free to take the 
microwave propagation vector $\bf k$
to lie in the $y-z$ plane.
With these definitions, after adding the 
most important electric-field-induced
terms, $\Lambda$ becomes\cite{note3:rwd}
\begin{eqnarray}
\fl 2\hbar \Lambda = - ({\rm b}_x+ \rmi {\rm b^m}_y)
\mu_0 \cos \Theta - 
\rmi(\epsilon_x+ \rmi \epsilon_y^m)
 d_W \Delta_{\beta_{-1/2} f_{-1/2}}
\nonumber \\
+ (E_x^u+\rmi E_y^u)
\epsilon_z d_2^2 \Delta_{\beta_{-1/2} e_{+1/2}}
- E_z^u
(\epsilon_x+\rmi \epsilon_y^m) d_2^2 \Delta_{\beta_{-1/2} f_{-1/2}}.
\label{eqc1}
\end{eqnarray}
Here, $d_W = \bar{\mathcal{V}}C_{1D}d_2$.
Under a polarization reversal $\phi 
\rightarrow - \phi$, the microwave components
$\epsilon_z$, $\epsilon_y^m$, and b$_x$ change
sign while $\epsilon_x$, b$_y^m$, and b$_z$ do
not.

The first two terms in Eq.~\ref{eqc1}
(with $\epsilon_y^m = b_y^m$ = 0)
are the amplitudes
A$_{\rm PC}$ and A$_{\rm PNC}$ discussed in the previous section
 and the last two terms
are the electric field induced amplitudes which are of concern
in this section. 
The first point is that
interferences among the electric field 
induced terms ($E1E1$ interferences)
are relatively small since there
are no applied electric fields. Also,
these terms are
readily eliminated because they are even
in the reversal of
$\bf k$, whereas the $E1M1$ PNC
pseudoscalar term is odd under
this reversal. Of the $E1M1$ 
interferences, those that
change sign under a reversal of the polarization 
$\phi$ are of most concern.
The two most significant effects are: 
(1) a term caused by a transverse
electric field
which we will
call the motional field effect, and,
(2) a term caused by an electric field
along $\bf B$ which we will call
the stray $E_z$ systematic effect.
In the following we estimate the size of these effects. 
Based on Eq.~\ref{eqb10} and Eq.~\ref{eqb6} and taking
$\tau=300 \mu$s and $\tan\phi$=0.1, 
we find:
\begin{equation}
\epsilon_x \rightarrow 0.16 V/cm, 
\epsilon_z \rightarrow 0.016 V/cm,
b_x \rightarrow 5 \times 10^{-5}G.
\label{fieldstrengths}
\end{equation}

\subsection{Motional electric field systematic effect}
The third term in 
Eq.~\ref{eqc1} arises from mixing of $\beta_{-1/2}$
with $e_{+1/2}$ by a transverse electric field and,
unlike the  PNC term ($2^{nd}$
term in Eq.~\ref{eqc1})
it is not resonant at the $\beta f$ crossing.
The largest term
is proportional to the combination
of fields, $\epsilon_z$ b$_y^m E_y^u$.
This involves an electric field transverse
to the magnetic field. The most significant
field of this kind arises from the
$\bm {v \times B}$ ``motional'' electric field. 

It is of interest to form the ratio $R^{mot}$ of the interference term
arising from the motional electric field to that of the 
pseudoscalar interference term. This is:
\begin{equation}
R^{mot}=8 \times 10^5
\frac{\langle E^{mot}\rangle \sin \chi }{C_{1D}},
\label{mot}
\end{equation}
where $\chi$ is the microwave misalignment angle defined by
b$_y^m$ = b$_x \sin \chi$, and $\langle E^{mot} \rangle$ is
the motional field (in Volts/cm)
averaged over the beam velocity distribution.
If the transverse velocity
distribution were cylindrically
symmetric and the axis of 
symmetry exactly coincident with a homogeneous magnetic
field, the motional field systematic would vanish. For
simplicity, we represent the imperfections of the real apparatus
by a misalignment between the beam velocity and the magnetic field.
Based on Fehrenbach's experiment\cite{feh93:dis} 
we take this angle to be 0.1 mrad. For a slow (77K)
deuterium beam,
and a magnetic field of $\sim$ 1200 G, we find
$\langle E^{mot} \rangle \sim 10^{-4} V/cm$.
Assuming the 
microwave misalignment angle $\sin \chi=10^{-4}$, we have:
\begin{equation}
R^{mot}\approx \frac {0.01} {C_{1D} }\approx 0.02
\label{sizermot}
\end{equation}
The motional field effect can be distinguished from the pseudoscalar term because it is
sensitive to slight changes in the direction of the B-field and it is non-resonant
at the crossing. 

For a new deuterium APV experiment, it would be desirable to
improve the
symmetry and alignment of the beam 
over what has been achieved in the past in order
to suppress this
effect by another order of 
magnitude so that it is
smaller than 0.3\% of the pseudoscalar
term.
But, regardless of the level of 
suppression obtained,
this effect can be accurately measured
and a correction can be applied to the final
result of the experiment.
To do this,
the experiment must incorporate
a continuous monitoring of the motional electric fields
 while the APV data are being taken. 
The general method for doing this is well
known, as it is an essential technique in APV experiments.
 The unwanted
fields are monitored by using
the atoms themselves
\cite{dun78:dis,gil86:pra}. 
In the present case, the motional field could
be monitored by, $e.g.$  
periodically moving the magnetic field
off the center of the level crossing
(retuning the microwave frequency to 
an adjacent resonator mode).
Away from the crossing, $\rm A_{PNC}$ is suppressed,
so the signal is dominated by the
motional electric field effect. The
limit on the motional field effect is
determined by the statistical accuracy of the
on-line monitoring, which by the arguments
of Section~\ref{sensitivity}, again reduces to
the need for an intense, slow metastable beam.

%which means this effect comes in at about 
%the level of precision for an
%``interesting'' measurement and so,
%although it would
%require considerable attention, 
%it does not rule out a new hydrogen
%parity experiment.
The largest systematic error in the measurement
of Fehrenbach was caused by  a motional electric field
and it is natural to ask why this is less problematic
here.
The first point is that the velocity of the slow D
beam is a factor of $\sim 300$ smaller than his 
fast beam so
the motional field is smaller by the same
factor. Also,
his experiment was performed at the
crossing of the $e$ and $\beta$ levels which
are mixed by the (transverse) motional field, so
for an experiment done
at the $\beta f$ crossing
there is 
an additional
relative suppression of approximately 10. The net result
is a suppression of a factor of
approximately 3000 for
the motional field effects in the experiment
discussed in Section~\ref{sensitivity}, 
relative to those in Fehrenbach's experiment.

\subsection{Stray $E_z$ systematic effect}
The last term in Eq.~\ref{eqc1} arises from mixing between
the $\beta_{-1/2}$ and $f_{-1/2}$ states by an
unwanted ``stray''
electric field parallel to the 
magnetic field. Such a field could
arise from surfaces 
in the interaction region, 
charged particles in the beam, $etc$.
We call the most significant term, proportional
to the fields $\epsilon_x$ b$_x E_z^u$,
the stray $E_z$ systematic effect.
The microwave field combination is
the same as that for the pseudoscalar
interference,
but this term has a dispersive shape
as a function of magnetic field. It
goes to zero near the center of the
$\beta_{-1/2} f_{-1/2}$ crossing, and has
a maximum at $\omega_{\beta_{-1/2} f_{-1/2}} \sim 
\gamma_{2p}/2$. The exact magnetic field
where the stray $E_z$ term vanishes depends
on couplings with more distant levels and it
can be calculated to high precision. 
Due to field inhomogeneities, frequency
shifts, the finite linewidth
of the transition, $etc$., one cannot
completely zero the stray $E_z$ effect. 
Rather, we
introduce a parameter $\delta_{\Delta B}$
which is the degree of suppression
of this term from its maximum value
obtained by setting the magnetic
field to the proper value.
There is an additional suppression of
this term since it is odd
under a reversal of $\bm B$, while
the pseudoscalar is even in this
reversal. We assign another
factor $\xi_B$ to account for this
suppression.

The ratio $R^{E_z}$ of the stray $E_z$
interference term to the pseudoscalar
is found using the same 
microwave field strengths
as previously:
\begin{equation}
R^{E_z} = 10^8 \frac
{\langle E_z^s \rangle \delta_{\Delta B} \xi_B}
{C_{1D}}
\label{strayEz}
\end{equation}
$\langle E_z^s \rangle$ is the average
electric field along z seen by the
atoms, expressed in V/cm.

The field $\langle E_z^s \rangle$
can be minimized by careful design of the interaction
region, but here we
just assume the
value achieved by Fehrenbach which was
$\langle E_z \rangle \sim 1 mV/cm$. We 
take the suppression factors to be
$\xi_B = \delta_{\Delta B} = 10^{-4}$,
based on his results. Using
these values we find:
\begin{equation}
R^{E_z} \approx \frac{10^{-3} } {C_{1D}}
\approx-2 \times 10^{-3}.
\label{sizestrayEz}
\end{equation}
where the coupling constant
was taken from Table~\ref{crossings}.
This estimate is an order of magnitude smaller than
the motional field effect, and is smaller than
our goal of 0.3\%
uncertainty in a measurement of the pseudoscalar term.
The stray $E_z$ systematic effect is readily measured 
while the APV data are being recorded
by setting the 
magnetic field off the crossing and looking for
a term which is polarization dependent, 
odd in the magnetic field,
and switches sign when the
field is tuned to the other side of the crossing.

\section{Summary\label{summary}}
The major argument for consideration of parity experiments
in H and D, that there is negligible atomic physics
uncertainty in the interpretation of the results, 
remains compelling. 
Parity violation experiments in deuterium could
play a role in looking for new physics by
providing a low energy measurement of 
sin$^2 \theta_W$ to confirm the SM
prediction 
for the  ``running of sin$^2 \theta_W$''.
A 0.3\% measurement of $C_{1D}$
would improve the limit on
sin$^2 \theta_W$ provided by APV experiments by a factor
of three. 

Hydrogen and deuterium
APV experiments could also contribute to
determination of the electron-nucleon weak
neutral current coupling constants 
listed in Table~\ref{limits}.
Only C$_{1D}$ is accurately constrained
by current data.
The uncertainty in C$_{1p}$ is about
the size of the SM prediction,
and C$_{2p}$ and
C$_{2D}$ are completely undetermined.
The values of the individual constants
could be determined by measuring APV at
several level crossings. An 
updated list of the SM predictions
for the parity violating matrix
elements at level crossings in
H and D are given in Table~\ref{crossings}.

In Section~\ref{sensitivity} we analyzed
a generic deuterium APV experiment, and
found that a 0.3\% measurement of 
C$_{1D}$ 
would be possible if a slow metastable
beam with an intensity of $\sim$ 5$\times 10^{14}$
deuterium atoms/s per hyperfine level 
were available. In 
\ref{secopticalpumping}
we discuss some ideas for obtaining
such a beam. 

In Section~\ref{systematics}, 
we discussed two systematic
effects and
found that, although they
provide a challenge for
a deuterium APV experiment, they
are not insurmountable obstacles.
After designing the experiment to minimize the 
systematic effects due to unwanted electric fields as
much as possible, the elimination of these effects
is done by monitoring them during the
APV data taking, using the atoms themselves 
to measure the fields. Thus, the limitation
on the experimental sensitivity due
these systematic effects is dependent on 
counting statistics, and, in this sense, also 
depends on the intensity of the metastable beam.

\ack
We are indebted to Zheng-Tian Lu for 
significant contributions and encouragement.
We also thank D. H. Beck for many useful discussions.
Work supported by 
The U.S. Department of Energy, Office of Nuclear Physics,
under Contract no. DE-AC02-06CH11357 (RJH).

\appendix
\section{Metastable quenching from beam-beam collisions
\label{secquenching} }

Metastable quenching from beam-beam
collisions places a practical
limit on
the density of metastable D(2s)
atoms in a beam suitable for
a deuterium APV experiment.
Forrey {\em et al.} \cite{for00.prl}
calculated the cross sections for ionization, excitation-transfer,
and elastic scattering in
collisions between metastable hydrogen 2s atoms at
thermal energies. The most important metastable
destruction mechanism was found to be double excitation
transfer to H(2p) :
\begin{equation}
H(2s)+H(2s)\rightarrow H(2p)+H(2p)
\label{beambeam}
\end{equation} 
for which they found a cross section $\sigma^{sp}=
9\times 10^{-12} cm^2$ at $E=4.1$ meV, varying as $E^{-1/2}$
at higher energies.

If we require that the mean free path $\lambda_{mfp}$ of
the D(2s) atoms be a meter, this
puts an upper limit on the metastable
density:
\begin{equation}
\rho_{2s} < \frac{1}{\sigma^{sp} \lambda_{mfp}}\approx
10^9 cm^{-3}.
\label{density}
\end{equation}
Then, for a beam 
with an average velocity of
10$^5$ cm/s, the D(2s) flux density
must be $<10^{14}$ cm$^{-2}$s$^{-1}$.
We can take this limit to apply to an individual
hyperfine component since one can quench out
all unwanted components immediately after
formation of the metastable beam using a 
deuterium spin filter\cite{mck68:prl}.
For a metastable beam of 5$\times$10$^{14}s^{-1}$ per hyperfine
component, the beam diameter
has to be $>$ 2.5 cm.

The metastable beam can also be quenched from beam-beam
collisions with D(1s) and
D$_2$. The cross section\cite{rya77:pra} is about 2 $\times$ $10^{-14}$cm$^2$,
so for a mean free path of 1 m at 10$^5$ cm/s, the flux density 
must be  $< 5 \times 10^{16}$cm$^{-2}s^{-1}$. For a 2.5 cm diameter
beam this corresponds to $2 \times 10^{17}$ D(1s)/s.

\section{Optical production of metastable hydrogen and deuterium beams}
\label{secopticalpumping}

One possibility for obtaining a slow metastable 
D(2s) beam with sufficient
intensity for a deuterium APV experiment is to use optical pumping 
starting with an intense dissociated deuterium beam. One such beam was produced 
by Harvey \cite{har82:jap}. In his scheme, 
a beam of hydrogen atoms in the ground state was pumped to
3p using a Lyman-$\beta$ discharge lamp. The 3p level decays to
the 2s state about 12\% of the time. This work produced 
a beam with a flux of only $2 \times 10^{6}$ metastable atoms/s because
of the low intensity of Lyman-$\beta$ radiation from the lamp.
The development of intense beams of UV radiation using
FEL technology might provide a considerable improvement in this
respect.

The photon absorption cross section\cite{bud04:ap}
for the 1s $\rightarrow$ 3p assuming the radiation
does not resolve the 3p fine structure is given by 
\begin{equation}
\sigma_{3p}=
       \frac{3 \lambda^2}{2\pi}\frac{\gamma_{3p1s}}{\gamma_{\rm line}},
\label{sigab}
\end{equation}
where $\lambda$ is the transition wavelength, $\gamma_{3p1s}$ 
is the ground state decay rate for the 3p level and 
$\gamma_{\rm line}$ is the transition linewidth.
We find:
\begin{equation}
\sigma_{1s3p}\approx \frac{1.3 \times 10^{-12}}{
\gamma_{\rm line}(GHz)}{\rm cm}^{2}.
\label{sigfel1s3p}
\end{equation}

We are interested in the amount of UV power at 102.6 nm needed to produce a metastable
D(2s) beam (in a single hyperfine component)
of 5 $\times 10^{14}$ D(2s) s$^{-1}$ to serve as a benchmark for 
FEL development. Assuming a ground state beam of 2 $\times$ $ 10^{17} s^{-1}$ with
a diameter of 2.5 cm and a UV beam with a 3 cm diameter and accounting for 
the 12$\%$ branching ratio for 3p $\rightarrow$ 2s, we find the power per bandwidth
required is 40 mW/GHz.

If intense Lyman-$\alpha$ radiation is
available, metastable
hydrogen can be obtained by a three step process;
two photon pumping 1s $\rightarrow$ 2p 
$\rightarrow$ 4d followed by decay to 2s. 
Lyman-$\alpha$ radiation drives the transition 
1s $\rightarrow$ 2p which is followed by a 2p $\rightarrow$ 4d 
transition saturated by a visible laser. Atoms in 4d have a
3\% branch to the 2s state\cite{note6:rwd}. This
scheme is similar to that used by Young, {\em et al.}
for production of metastable Kr atoms \cite{you02:jpb,din07:rsi}.
Although the branching ratio for 
the 4d state is not as favorable as
for the 3d state, the oscillator strength for the
transition 1s $\rightarrow$ 2p ($f_{ab}$ = 0.416) is five times
larger than
for the 1s $\rightarrow$ 3p transition. Also, an
FEL capable of producing several watts of Lyman-$\alpha$
radiation is under development at Jefferson Lab in
Virginia\cite{gpw06:nt1}. 

Another
possibility for production of a metastable H beam is  
based on the two-photon
transition 1s $\rightarrow$ 2s with 243 nm radiation
 \cite{bas77:prl,bos89:pra,dor97:pra,yat99:pra}.
This is a weak process since it does not involve an
intermediate resonance state, and requires high
intensity radiation. 
The needed intensity is normally obtained by
focussing
the light to a small spot, but this is 
incompatible with the limitation on
the density of the metastable hydrogen atoms discussed
in \ref{secquenching}, which requires 
a photon spot size of at least 5 cm$^2$. 
For this reason, two-photon excitation does
not appear to be a viable option for production
of the required metastable beam. This conclusion 
is also supported by existing experimental 
data \cite{lu01:bap,bec07:pc}.

\section{Normalization Transition\label{norm}}

In the generic experiment
discussed in Section~\ref{sensitivity},
the psuedoscalar
interference 
term 
is measured by picking out the
part of the $\alpha_{+1/2} \rightarrow
\beta_{-1/2}$ transition rate that changes sign under the 
polarization reversal
$\phi \rightarrow -\phi$. 
Ideally, the microwave field b$_x$ that
drives the PC transition 
would have the same strength in each
polarization state, so the PC part
of the transition would average to zero
over the course of the experiment.
In practice, it is not possible to achieve this
to sufficient accuracy, and several
independent reversals are used\cite{lev84:pra}.
However, it is possible to monitor
the field
strength in each polarization state during
accumulation of APV data, and
make this reversal exact within the
statistical uncertainty of the
monitor.
One way to do this is by 
periodically switching on an auxiliary 
static electric field $\bm E^N$ perpendicular to $\bf B$.
This adds a term to Eq.~\ref{eqb3}
of the form:
\begin{eqnarray}
2\hbar \Lambda_N = d^2_2 \Delta_{\beta_{-1/2} e_{+1/2}} (E_x^{(N)} + \rmi E_y^{(N)})
\epsilon_z 
\label{eqn1}
\end{eqnarray} 
The field direction can be chosen so that the new term
is nearly orthogonal to the PNC amplitude.
$\Lambda_N$ is proportional to the
microwave electric field component $\epsilon_z$
which is directly proportional 
to the microwave magnetic field 
component b$_x$ that
needs to be measured.
(See Fig.~1).
The field strength $E^{(N)}$ is chosen so
that $\Lambda_N$
is much larger than $\Lambda$
(Eq.~\ref{eqb3}) 
so that, when $\bm E^{(N)}$ is on, $\Lambda_N$
dominates
the $\alpha_{+1/2} 
\rightarrow \beta_{-1/2}$ transition
rate, and the
contribution of the PNC interference
term is sufficiently small that it does
not contribute to the measurement of the
 field strength.

%Technical detail that it is necessary to include, but that interrupts the flow 
%of the article, may be consigned to an appendix.  Any appendices should be 
%included at the end of the main text of the paper, after the acknowledgments 
%section (if any) but before the reference list.  If there are two or more 
%appendices they should be called Appendix A, Appendix B, etc.  Numbered 
%equations will be in the form (A.1), (A.2), etc, figures will appear as figure 
%A1, figure B1, etc and tables as table A1, table B1, etc.
%The command \verb"\appendix" is used to signify the start of the appendices. 
%Thereafter \verb"\section", \verb"\subsection", etc, will give headings 
%appropriate for an appendix. To obtain a simple heading of `Appendix' use the 
%code \verb"\section*{Appendix}". If it contains numbered equations, figures or 
%tables the command \verb"\appendix" should precede it and 
%\verb"\setcounter{section}{1}" must follow it. 

\section*{References}
%when you are ready to submit, comment out this next line and past the *.bbl 
%file in here
%\bibliography{hpnc06}

\begin{thebibliography}{10}
\bibitem{note7:rwd}
An overall factor of $\cos \Theta_+ \cos \Theta_-$ has been left out of
  Eq.~\ref{eqb3}. Here, $\Theta_{\pm}$ are hyperfine mixing parameters for
  deuterium\cite{dun78:dis,dun78:pra}. At the $\beta$f crossing $\cos \Theta_+$
  $\simeq$ $\cos \Theta_-$ $\simeq$ 0.999.
\bibitem{note3:rwd}
For the purposes of this article, it is sufficient to include only the mixings
  of $\beta_{-1/2}$ with the 2{P}$_{1/2}$ levels. We neglect small hyperfine
  induced mixings, perturbations of $\alpha_{+1/2}$ by electric fields, and
  mixings with more distant p states. All these effects need to be considered
  in a full treatment.
\bibitem{note6:rwd}
Another possibility would be to drive $2p \rightarrow 4s$ in the second step.
  The $4s$ state has a larger branch to 2s (5\%), but the oscillator strength
  for this transition, is a factor of 40 smaller than that for $2p \rightarrow
  4d$.
\end{thebibliography}
%\end{document}

\end{document}